\documentclass[useAMS,usenatbib,usegraphicx]{mn2e}

\usepackage{amsmath,amssymb,bm,upgreek,color}
\usepackage{mathrsfs}
\usepackage[latin1]{inputenc}
\usepackage{JournalNames}
\usepackage{upgreek}
\usepackage{enumerate}
\usepackage{wrapfig}
\usepackage{multicol}
\usepackage{lscape}
\usepackage{pifont}
\usepackage{subfigure}

\begin{document}

\title[Acceleration of the universe: a reconstruction ....]
{Acceleration of the universe: a reconstruction of the effective equation of state}

\author[Ankan Mukherjee]{Ankan Mukherjee  \\
 \it Department of Physical Sciences, Indian Institute of Science Education and Research Kolkata, Mohanpur, West Bengal-741246, India}

\date{Accepted ??. Received ??; in original form ??}

\pagerange{\pageref{firstpage}--\pageref{lastpage}} \pubyear{????}

\maketitle

\label{firstpage}

\newcommand{\be}{\begin{equation}}
\newcommand{\ee}{\end{equation}}
\newcommand{\bea}{\begin{eqnarray}}
\newcommand{\eea}{\end{eqnarray}}

\begin{abstract}
The present work is based upon a parametric reconstruction of the effective or total equation of state in a model for the universe with accelerated expansion. The constraints on the model parameters are obtained by maximum likelihood analysis using the supernova distance modulus data, observational Hubble data, baryon acoustic oscillation data and cosmic microwave background shift parameter data. For statistical comparison, the same analysis has also been carried out for  the $w$CDM dark energy model. Different model selection criteria (Akaike information criterion (AIC)) and  (Bayesian Information Criterion (BIC)) give the clear indication that the reconstructed  model is well consistent with the $w$CDM model. Then both the models ($w_{eff}(z)$ model and $w$CDM model) have also been presented through ($q_0$,$j_0$) parameter space.  Tighter constraint on the present values of dark energy equation of state parameter ($w_{DE}(z=0)$) and cosmological jerk ($j_0$) have been achieved for the reconstructed model. 
\end{abstract}


\section{Introduction} 
A new era in cosmological research has begun with the discovery of the late time cosmic acceleration. Two groups in the late nineties first observed this phenomenon while observing the type Ia supernovae \citep{Riess1998,Perlmutter1999}. Later on, observations by different groups confirmed the accelerated expansion of the universe \citep{Knop2003,Tonry2003,Barris2004,Hicken2009,Suzuki2012}. This has come out as the most puzzling phenomenon in cosmology as gravity, as we know from any local phenomenon, is attractive. Another important point to be noted is that the analysis of the observational data reveals the fact that the acceleration has started recently \citep{Riess2004}. 

\par Different theoretical prescriptions are there in the literature to explain the cosmic acceleration. One of these is the {\it dark energy} which is actually a hypothetical exotic component introduced in the matter sector of the universe. Dark energy, theoretically  constructed in such a way that it induces negative pressure, is the candidate responsible for the alleged acceleration. There are some excellent review articles where the theoretical framework of dark energy and different dark energy models have been comprehensively discussed \citep{SahniStaro2000,Carroll2001,PeeblesRatra2003,Padmanabhan2003,CopeSamiTsuji2006,Martin2008}. The most popular and well consistent with most of the observational data is the $\Lambda$CDM model where the constant vacuum energy density, dubbed as cosmological constant, serves as the dark energy candidate. But the cosmological constant suffers from the problem of fine tuning due to the humongous discrepancy between the observationally required value of cosmological constant and the theoretically calculated one. There are reviews on the cosmological constant model where those issues have been discussed in great details \citep{Carroll2001,Padmanabhan2003}.

\par The dark energy models are the attempts to explain cosmic acceleration where  General Relativity (GR) is employed as the correct theory of gravity and the matter sector includes some exotic component. Obviously the other option is to look for the modification of GR which allows the cosmic acceleration without introducing any exotic component. Scalar-tensor theories \citep{BertolamiMartins2000,BanerjeePaovn2001a,BanerjeePaovn2001b,SenSen2001,MotaBarrow2004a,MotaBarrow2004b,DasBanerjee2008}, $f(R)$ gravity theory \citep{CCT2003,CDTT2004,Vollick2003,NojiriOdintsov2003,Carroll2005,MenaSantiago2006,NojiriOdintsov2006,DasBanerjeeDadhich2006,NojiriOdintsov2007a,NojiriOdintsov2007b,NojiriOdintsov2009}, different higher dimensional gravity theories \citep{DeffayetDvaliGabadadze2002,NojiriOdintsovSami2006,DvaliGabadadzePorrati2008,HossainMSS2014,Bamba2014} etc are based upon the modification of GR.

\par Though different theoretical approaches are there to explain the phenomenon of cosmic acceleration, till now none of them is definitely known as the appropriate one. The present trend of modelling of late time cosmic acceleration is called {\it reconstruction} where the model is built up by taking the observational data directly into account. This is actually the {\it reverse way} of finding the suitable cosmological model. This type of attempt to find the scalar field potential has been discussed long ago by Ellis and Madsen \citep{EllisMadsen1991}. There are two types of reconstruction, parametric and non-parametric. The parametric reconstruction  is based upon the estimation of the model parameters from different observational data. It is also called the model dependent approach. The prime idea is to assume a particular evolution scenario and then to find the nature of the matter sector or the exotic component which is responsible for the alleged acceleration. In the context of dark energy, this method was first discussed by Starobinsky \citep{Starobinsky1998} where the density perturbation has been used in the context of reconstruction. Data of cosmological distance measurement has been invoked in the context of reconstruction by Huterer and Turner \citep{HutererTurner1999,HutererTurner2001} and also by Saini {\it et al} \citep{SainiRSS2000}. Some other earlier works on parametric reconstruction have been referred to in \citep{ChevallierPolarski2001,Linder2003,CoorayHuterer1999,MaorBrusteinSteinhardt2001,WellerAlbrecht2001,GerkeEfstathiou2002,GongWang2007}. Recently different parametrization of dark energy equation of state have been explored by Xia, Li and Zhang \citep{XiaLiZhang20013} and also by Hazra {\it et al.} \citep{HazraDK2015} where the most recent observational data sets have been adopted. The second kind of reconstruction, the non-parametric one, is based upon rigorous statistical analysis  of  the observational data rather than any prior assumption of the parametric from of any cosmological parameter. The primary endeavour of non-parametric reconstruction is to find the nature of cosmic evolution directly from observational data      \citep{SLP2005,SLP2007,Holsclaw2010,Holsclaw2011,Crittenden2012,NairJJ2014}.  

\par Kinematic approach in the  study of cosmic evolution is independent of any particular gravity theory. The deceleration parameter, the jerk parameter etc belong to the set of kinematic quantities. Reconstruction of different kinematic quantities using the observational data  depict the nature of cosmic evolution without presuming anything about dark energy or any particular gravity theory.  A  kinematic approach  was discussed by Riess {\it et al.} \citep{Riess2004}, where a linear parametrization of deceleration parameter $q(z)$ has been used to estimate the value of redshift at which the transition from decelerated to accelerated expansion happened. The cosmological jerk parameter, which is a dimensionless representation of the 3rd order time derivative of the scale factor, has been used as a diagnostic of dark energy models by Sahni {\it et al} \citep{SSSAlam2003} and Alam {\it et al} \citep{AlamSSS2003}.  The reconstruction of jerk from future data was also indicated by Sahni {\it et al} \citep{SSSAlam2003} and Alam {\it et al} \citep{AlamSSS2003}. There the jerk parameter and a combination of jerk and deceleration parameter together have been stated as the statefinder diagnostic. Reconstruction of dark energy equation of state through the parameterization of cosmological jerk has been discussed by Luongo \citep{Luongo2013}. Kinematic approach to the modelling of accelerating universe has been discussed by Rapetti {\it et al} \citep{Rapetti2007}, where a constant jerk parameter model has been invoked. Evolving jerk parameter models has been investigated by Zhai {\it et al} \citep{zz2013} and by Mukherjee and Banerjee \citep{MukherjeeBanerjee2016}.
 
\par In the present work, a parametric reconstruction of the effective or total equation of state has been presented. The functional form of  effective equation of state parameter is chosen in such a way that it tends to zero at high value of redshift which is the signature of matter dominated universe. The present value of the effective equation of state parameter depends on the model parameters which have been constrained from the observational data.  The constraints on the model parameters are obtained by $\chi^2$ minimization technique (which is equivalent to the maximum likelihood analysis) using different observational data sets. Here the distance modulus data of type Ia supernovae (SNe), observational Hubble data (OHD), baryon acoustic oscillation data (BAO) and cosmic microwave background (CMB) distance prior named the CMB shift parameter (CMBShift) have been adopted.

\par The present work is not based on a purely kinematical approach, it rather assumes GR as the theory of gravity, but there is hardly any prior assumption about the distribution of the components in the matter sector. A reconstruction of effective equation of state does not depend upon the individual properties of the different components of the matter sector. The possibility of interaction between the components can also be investigated in this case. The prime endeavour of this reconstruction is to figure out the distribution of the matter components instead of any prior assumption about them. For comparison, a standard dark energy model, the $w$CDM, has also been explored using the same data sets. There are some significant differences in the prior assumptions of the $w$CDM model and the reconstructed $w_{eff}$ model. For the $w$CDM model, the dark energy equation of state parameter ($w_{DE}$) is assumed to be a constant throughout the evolution. At the same time, the dark matter is allowed to have an independent conservation. The cosmological constant model or the $\Lambda$CDM and the $w$CDM model are at present the most popular dark energy models as they are well consistent with most of the observational data. For these reasons, the $w$CDM model has been chosen in the present work as an example for a comparison with the reconstructed model. Different mode selection criteria unambiguously  show the consistency of this model with the standard $w$CDM dark energy model. For direct comparison, both the models have also been presented through ($q_0$,$j_0$) parameter space, where $q_0$ is the present value of deceleration parameter and $j_0$ be the present value of jerk parameter.

\par The following sections contain the mathematical formulation of the reconstruction of effective equation of state (section 2), brief discussion about the observational data sets used for the statistical analysis (section 3) and the result of the statistical analysis along with the plots of  likelihood as a function of the parameters and the plots of confidence contours on 2D parameter space and  detail comparison with the $w$CDM dark energy model (section 4 and section 5). In section 6, an overall discussion regarding the reconstructed model and the results obtained have been discussed.

\section{Reconstruction of the model}

The basic mathematical framework of cosmology is the Friedmann model where the line element for a spatially flat universe  is written as 
\be
ds^2=-dt^2+a^2(t)\Big[dx^2+dy^2+dz^2\Big].
\ee
This is the well-known Friedmann-Robertson-Walker (FRW) metric for spatially flat geometry. Incorporating the FRW metric to Einstein's field equations, the Friedmann equations for spatially flat universe are obtained as 
\be
3H^2=8\pi G\rho,
\label{friedmann1}
\ee

\be
2\dot{H}+3H^2=-8\pi Gp,
\label{friedmann2}
\ee
where $H$ is the Hubble parameter defined as $H=\frac{\dot{a}}{a}$ (an over-headed dot denotes the derivative with respect to cosmic time $t$), the $\rho$ is the total energy density and the $p$ is the pressure. Now the effective or total equation of state parameter ($w_{eff}$) is defined as
\be
w_{eff}=\frac{p}{\rho}.
\ee
This $\rho$ and $p$ take care of the density and the pressure respectively for all the forms of the matter present in the universe. Now using equations (\ref{friedmann1}) and (\ref{friedmann2}), the effective equation of state parameter is written as
\be
w_{eff}=-\frac{2\dot{H}+3H^2}{3H^2}.
\label{weff}
\ee
 It is convenient to use redshift (z) as the argument instead of cosmic time $t$ as $z$ is a dimensionless quantity. If the argument of differentiation of $H$ is changed from cosmic time $t$ to redshift $z$. Redshift z is defined as $z+1=\frac{a_0}{a}$, where $a_0$ is the present value of the scale factor. Then the relation is given as
\be
\dot{H}=-(1+z)H\frac{dH}{dz}.
\label{dHdz}
\ee
Now one ansatz is required to close the system of equations, namely the equation (\ref{friedmann1}) and (\ref{friedmann2}). In the present work,  a parametric form of the effective equation of state $w_{eff}$ as a function of redshift $z$ is assumed as
\be
w_{eff}=-\frac{1}{1+\alpha(1+z)^n},
\label{weffz}
\ee
where $\alpha$ and $n$ are two model parameters. It is now clear from the observation of large scale structure and the existing models of structure formation that the contribution to the energy budget of the universe was dominated by dark matter at high redshift. At recent era, the prime contribution is coming from the exotic component dubbed as dark energy. As the dark matter is pressure less, the effective equation of state at high redshift was effectively zero. At the epoch of recent acceleration, it has a negative value which is less than $-\frac{1}{3}$. The functional form of the effective equation of state (equation (\ref{weffz})) assumed for the present reconstruction can easily accommodate these two phases of evolution. For positive values of the model parameter $\alpha$ and $n$, the values of $w_{eff}(z)$ tends to zero a high value of the redshift $z$ and at $z=0$, its value depends upon the upon the model parameter $\alpha$. It is also clear from the expression of $w_{eff}(z)$ (equation (\ref{weffz})) that a positive value of the model parameter $\alpha$ always fixes a lower bound to the value of $w_{eff}(z)$ and keeps it in the non-phantom regime.

\par Introducing the assumed ansatz of $w_{eff}(z)$ (equation (\ref{weffz})) to equation (\ref{weff}) and (\ref{dHdz}), the differential equation for $H$ reads as
\be
\frac{2}{3}(1+z)\frac{1}{H}\frac{dH}{dz}-1=-\frac{1}{1+\alpha(1+z)^n}.
\ee
And the solution obtained for the Hubble parameter as a function of redshift is 
\be
H(z)=H_0\Bigg(\frac{1+\alpha(1+z)^n}{1+\alpha}\Bigg)^{\frac{3}{2n}},
\label{hubbleparam}
\ee
where $H_0$ is the value of Hubble parameter at $z=0$. One interesting point regarding this expression of Hubble parameter is that for $n=3$, this becomes exactly like the $\Lambda$CDM model. Hence the estimated value of the model parameter $n$ will clearly indicate whether a $\Lambda$CDM or a time evolving dark energy is preferred by observations. 

\par It is imperative to note at this point that in the series expansion of $h^2(z)$ (where $h(z)=H(z)/H_0$), which can be obtained from equation (\ref{hubbleparam}), there will be a term with $(1+z)^3$. This corresponds to the dark matter density. The coefficient of $(1+z)^3$ is $\left(\frac{\alpha}{1+\alpha}\right)^{3/n}$. It is equivalent to the matter density parameter $\Omega_{m0}$ which is the ratio of present matter density and the present critical density ($3H_0^2/8\pi G$). Thus the contribution of the dark energy can be obtained by subtracting this term from $h^2(z)$,
\be 
\Omega_{DE}(z)=h^2(z)-\Big(\frac{\alpha}{1+\alpha}\Big)^{\frac{3}{n}}(1+z)^3,
\ee
($\Omega_{DE}$ is the dark energy density scaled by the present critical density). Similarly the pressure contribution of the dark energy can be obtained using equation (\ref{friedmann1}) and (\ref{friedmann2}) along with the expression of Hubble parameter $H(z)$ obtained in equation (\ref{hubbleparam}) . Finally the dark energy equation of state parameter can be written as a function of redshift and the associated model parameters as

\be
{\small
w_{DE}(z)= -\frac{\Bigg(\frac{1+\alpha(1+z)^n}{1+\alpha}\Bigg)^{\frac{3}{n}}-\Big(\frac{\alpha}{1+\alpha}\Big)(1+z)^n\Bigg(\frac{1+\alpha(1+z)^n}{1+\alpha}\Bigg)^{\frac{3}{n}-1}}{\Bigg(\frac{1+\alpha(1+z)^n}{1+\alpha}\Bigg)^{\frac{3}{n}}-\Big(\frac{\alpha}{1+\alpha}\Big)^{\frac{3}{n}}(1+z)^3}.
}
\ee
It is clear from the expression of $w_{DE}(z)$ that for $n=3$, the value $w_{DE}=-1$, which is the $\Lambda$CDM.

\section{Observational Data}
Four observational data sets have been used for the statistical analysis of the model in the present work. These are the observational Hubble data (OHD), distance modulus data from type Ia supernove (SNe), baryon acoustic oscillation (BAO) data along with the value of acoustic scale at photon electron decoupling and the ratio of comoving sound horizon at decoupling and at drag epoch estimated from CMB radiation power spectrum and the CMB shift parameter (CMBShift) data. The discussion about the observational data has also been presented in a very similar fashion by Mukherjee and Banerjee \citep{MukherjeeBanerjee2016}.

\begin{table}
\caption{{\small $H(z)$ data table (in unit [$km$ $s^{-1}$$Mpc^{-1}$])
}}
\begin{center}
\resizebox{0.37 \textwidth}{!}{  
\begin{tabular}{ c |c |c c c } 
 \hline
 \hline
 $z$  & H & $\sigma_H$ & References \\ 

 \hline
0.07      & 69      & 19.6  & \cite{Zhang2014}\\ 
0.1       & 69      & 12    & \cite{Simon2005}\\ 
0.12      & 68.6    & 26.2  & \cite{Zhang2014}\\ 
0.17      & 83      & 8     & \cite{Simon2005}\\ 
0.179     & 75      & 4     & \cite{Moresco2012}\\ 
0.199     & 75      & 5     & \cite{Moresco2012}\\ 
0.2       & 72.9    & 29.6  & \cite{Zhang2014}\\ 
0.27      & 77      & 14    & \cite{Simon2005}\\ 
0.28      & 88.8    & 36.6  & \cite{Zhang2014}\\ 
0.35      & 76.3    & 5.6   & \cite{ChuangWang2013}\\ 
0.352     & 83      & 14    & \cite{Moresco2012}\\ 
0.4       & 95      & 17    & \cite{Simon2005}\\ 
0.44      & 82.6    & 7.8   & \cite{Blake2012}\\ 
0.48      & 97      & 62    & \cite{Stern2010}\\ 
0.593     & 104     & 13    & \cite{Moresco2012}\\ 
0.6       & 87.9    & 6.1   & \cite{Blake2012}\\ 
0.68      & 92      & 8     & \cite{Moresco2012}\\ 
0.73      & 97.3    & 7     & \cite{Blake2012}\\ 
0.781     & 105     & 12    & \cite{Moresco2012}\\ 
0.875     & 125     & 17    & \cite{Moresco2012}\\ 
0.88      & 90      & 40    & \cite{Stern2010}\\ 
0.9       & 117     & 23    & \cite{Simon2005}\\ 
1.037     & 154     & 20    & \cite{Moresco2012}\\ 
1.3       & 168     & 17    & \cite{Simon2005}\\ 
1.43      & 177     & 18    & \cite{Simon2005}\\ 
1.53      & 140     & 14    & \cite{Simon2005}\\ 
1.75      & 202     & 40    & \cite{Simon2005}\\ 
2.34      & 222     & 7     & \cite{Delubac2015}\\

 \hline
 \hline
\end{tabular}
}
\end{center}
\label{hubbtable}
\end{table}

\subsection{Observational Hubble parameter data:}
Here the measurement of Hubble parameter $H(z)$ by different groups have been used. The estimation of the value of $H(z)$ can be obtained from the measurement of differential of redshift $z$ with  respect to cosmic time $t$ as
\be
H(z)=-\frac{1}{(1+z)}\frac{dz}{dt}.
\label{Hz}
\ee
The differential age of galaxies have been used as an estimator of $dz/dt$ by Simon {\it et al.}  \citep{Simon2005}. Measurement of cosmic expansion history using red-enveloped galaxies was done by Stern {\it et al} \citep{Stern2010} and by Chuang and Wang \citep{ChuangWang2013}. Measurement of expansion history from WiggleZ Dark Energy Survey has been discussed by Blake {\it et al.} \citep{Blake2012}. Measurement of Hubble parameter at low redshift  using the differential age method along with Sloan Digital Sky Survey (SDSS) data have been presented by Zhang {\it et al}  \citep{Zhang2014}. Compilation of observational Hubble parameter measurement has been presented by Moresco {\it et al} \citep{Moresco2012}. Finally, the measurement of Hubble parameter at $z=2.34$ by Delubac {\it et al} \citep{Delubac2015} has also been used in the present analysis. Table \ref{hubbtable} presents the $H(z)$ measurements which have been adopted in the present analysis. The measurement of $H_0$ from Planck+lensing+WP+lightL \citep{AdePlanck2014} has also been used in the analysis. The  model parameter values can be estimated using $\chi^2$-statistics, defined as
\be
\chi^2_{{\tiny OHD}}=\sum_{i}\frac{[H_{obs}(z_i)-H_{th}(z_i,\{\theta\})]^2}{\sigma_i^2},
\label{chiOHD}
\ee 
where $H_{obs}$ is the observed value of the Hubble parameter, $H_{th}$ is theoretical one and $\sigma_i$ is the uncertainty  associated to the $i^{th}$ measurement. And the $\chi^2$ is a function of the set of  model parameters $\{\theta\}$.

\begin{figure*}
\begin{center}
\includegraphics[angle=0, width=0.22\textwidth]{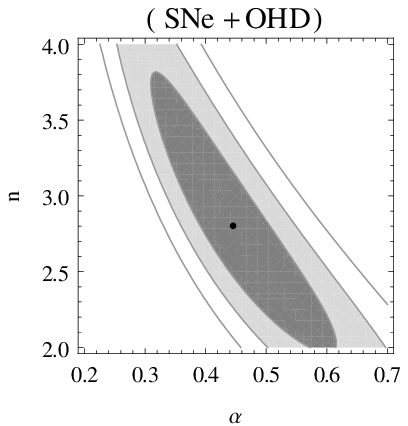}
\includegraphics[angle=0, width=0.22\textwidth]{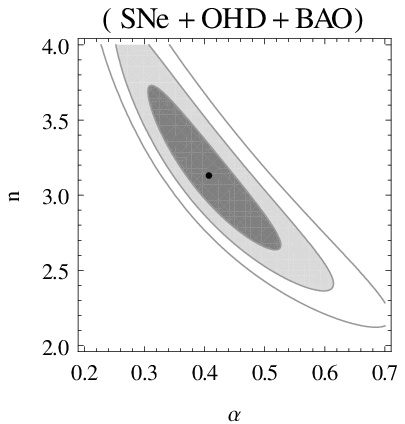}
\includegraphics[angle=0, width=0.215\textwidth]{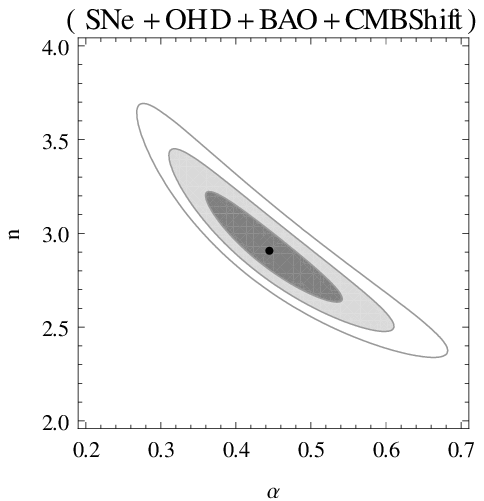}
\end{center}
\caption{{\small Confidence contours on the 2D parameter space of the $w_{eff}(z)$ model obtained for obtained for different combinations of the data sets. 1$\sigma$, 2$\sigma$ and 3$\sigma$ confidence regions are presented from inner to outer portion and the central black dots represent the corresponding  best fit points. The left panel shows the confidence contours obtained for SNe+OHD, the middle panel shows the confidence contours obtained for SNe+OHD+BAO and the right panel shows confidence contours for SNe+OHD+BAO+CMBShift.}}
\label{weffcontour}
\end{figure*}
\begin{figure*}
\begin{center}
\includegraphics[angle=0, width=0.23\textwidth]{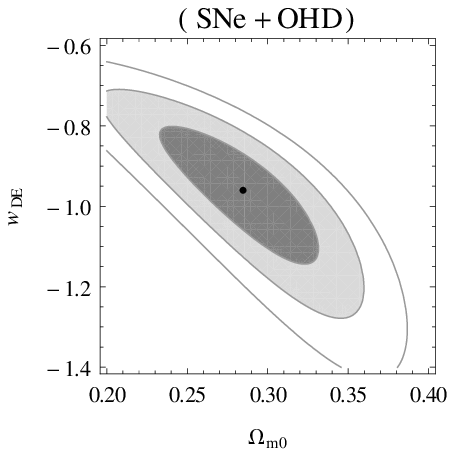}
\includegraphics[angle=0, width=0.23\textwidth]{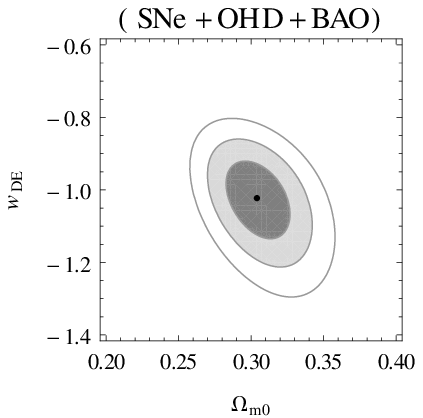}
\includegraphics[angle=0, width=0.226\textwidth]{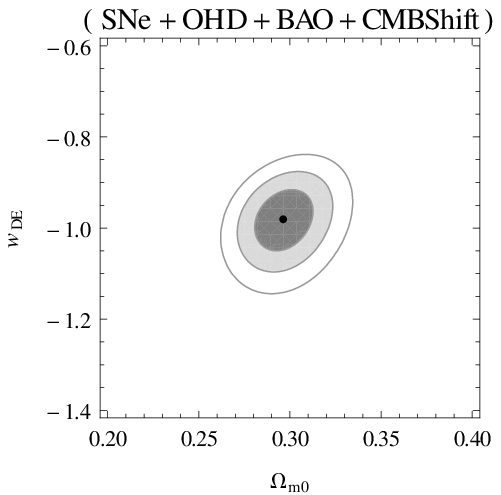}
\end{center}
\caption{{\small Confidence contours on the 2D parameter space of the $w$CDM model obtained for obtained for different combinations of the data sets. 1$\sigma$, 2$\sigma$ and 3$\sigma$ confidence regions are presented from inner to outer portion and the central black dots represent the corresponding  best fit points. The left panel shows the confidence contours obtained for SNe+OHD, the middle panel shows the confidence contours obtained for SNe+OHD+BAO and the right panel shows confidence contours for SNe+OHD+BAO+CMBShift.}}
\label{wCDMcontour}
\end{figure*}

\subsection{Type Ia supernova data:} 
The data from supernova observations is the most widely used data sample to study the late time dynamics of the universe. The distance modulus of type Ia supernova is the difference between the apparent magnitude ($m_B$) and absolute magnitude ($M_B$) of the B-band of the observed spectrum. It is defined as
\be
\mu(z)=5\log_{10}{\Bigg(\frac{d_L(z)}{1Mpc}\Bigg)}+25,
\ee
where the $d_L(z)$ is the luminosity distance and in a spatially flat FRW universe it is defined as
\be
d_L(z)=(1+z)\int_0^z\frac{dz'}{H(z')}.
\ee
In the present work, the 31 binned distance modulus data sample of the recent joint lightcurve analysis (jla) \citep{Betoule2014} has been utilized. To account for the correlation between different bins, the formalism discussed by Farooq, Mania and Ratra \citep{FarooqManiaRatra2013} has been adopted. The $\chi_{SNe}^2$ has been defined as
\be
\chi_{SNe}^2=A(\{\theta\})-\frac{B^2(\{\theta\})}{C}-\frac{2\ln{10}}{5C}B(\{\theta\})-Q,
\ee     
where 
\be 
A(\{\theta\})=\sum_{\alpha,\beta}(\mu_{th}-\mu_{obs})_{\alpha}(Cov)^{-1}_{\alpha\beta}(\mu_{th}-\mu_{obs})_{\beta},
\ee
\be
B(\{\theta\})=\sum_{\alpha}(\mu_{th}-\mu_{obs})_{\alpha}\sum_{\beta}(Cov)^{-1}_{\alpha\beta},
\ee
\be
C=\sum_{\alpha,\beta}(Cov)^{-1}_{\alpha\beta},
\ee
and the $Cov$ is the $31\times31$ covarience matrix of the binned data. Here the $Q$ is a constant which does not depends upon the parameters and hence has been ignore.

\subsection{Baryon acoustic oscillation data:}
In the present work, the Baryon acoustic oscillation (BAO) data along with the {\it acoustic scale ($l_A$)}, the {\it comoving sound horizon ($r_s$)} at photon decoupling epoch ($z_*$) and at drag epoch ($z_d$) as measured by Planck \citep{AdePlanck2014,WangWang2013b} has been utilized. The comoving sound horizon at photon decoupling is defined as
\be
r_s(z_*)=\frac{c}{\sqrt{3}}\int_0^{1/(1+z_*)}\frac{da}{a^2H(a)\sqrt{1+a(3\Omega_{b0}/4\Omega_{\gamma 0})}},
\label{rs}
\ee
where $\Omega_{b0}$ is the present value of Baryon density parametrer and $\Omega_{\gamma 0}$ is the present value of photon density parameter. According to the Planck results, the value of redshift at photon decoupling is $z_*\approx 1091$ and reshift at drag epoch is $z_d\approx 1021$ \citep{AdePlanck2014}.  The acoustic scale at decoupling is defined as
\be
l_A=\pi\frac{d_A(z_*)}{r_s(z_*)},
\label{accscale}
\ee
where $d_A(z_*)=c\int_0^{z_*} \frac{dz'}{H(z')}$, the {\it comoving angular diameter distance} at decoupling. Another important definition  is of {\it dilation scale} $D_V(z)=[czd_A^2(z)/H(z)]^{\frac{1}{3}}$. Here we have taken three mutually uncorrelated measurements of $\frac{r_s(z_d)}{D_V(z)}$, the result of 6dF Galax Survey at redshift $z=0.106$ \citep{Beutler2011}, and the results of Baryon Oscillation Spectroscopic Survey (BOSS) at redshift $z=0.32$ (BOSS LOWZ) and at redshift $z=0.57$ (BOSS CMASS) \citep{Anderson2014}. The Planck measurements of {\it acoustic scale ($l_A$)} and the ratio of {\it comoving sound horizon ($r_s$)} at two different epoch (the drag eopch ($z_d$) and at decoupling epoch ($z_*$)), are given as $l_A=301.74\pm0.19$, $\frac{r_s(z_d)}{r_s(z_*)}=1.019\pm0.009$  \citep{AdePlanck2014,WangWang2013b}. Finally the ratio $\Big(\frac{d_A(z_*)}{D_V(z_{BAO})}\Big)$ at three different values of $z_{BAO}$ have been obtained combining the Planck results with the BAO measurements. These can be used to obtain the BAO/CMB constraints on dark energy models. Table \ref{tableBAO} contains the values of $\Big(\frac{r_s(z_d)}{D_V(z_{BAO})}\Big)$ and finally the $\Big(\frac{d_A(z_*)}{D_V(z_{BAO})}\Big)$ at three different redshift of BAO measurement.

\begin{table*}
\caption{\small BAO/CMB data table}.
\begin{center}
\resizebox{0.6\textwidth}{!}{  
\begin{tabular}{ |c ||c |c |c |} 
 \hline
  \hline
 $z_{BAO}$ & 0.106 & 0.32 & 0.57 \\ 
 \hline
 \hline
 $\frac{r_s(z_d)}{D_V(z_{BAO})}$ & 0.3228$\pm$0.0205 & 0.1167$\pm$0.0028 & 0.0718$\pm$0.0010 \\ 
 \hline
 $\frac{d_A(z_*)}{D_V(z_{BAO})}\frac{r_s(z_d)}{r_s(z_{*})}$ & 31.01$\pm$1.99 & 11.21$\pm$0.28  & 6.90$\pm$0.10 \\ 
 \hline
 $\frac{d_A(z_*)}{D_V(z_{BAO})}$ & 30.43$\pm$2.22 & 11.00$\pm$0.37  & 6.77$\pm$0.16\\ 
 \hline

\end{tabular}
}
\end{center}
\label{tableBAO}
\end{table*}

The relevant $\chi^2$, namely $\chi^2_{BAO}$, is defined as:
\be
\chi^2_{BAO}={\bf X^{t}C^{-1}X},
\label{chibao}
\ee
where
\[
{\bf X}=
\left( {\begin{array}{c}
   \frac{d_A(z_*)}{D_V(0.106)}-30.43   \\
   \frac{d_A(z_*)}{D_V(0.2)}-11.00     \\
   \frac{d_A(z_*)}{D_V(0.35)}-6.77  \\
\end{array} } \right)
\]
and $C^{-1}$ is the inverse of the covariance matrix. As the three measurements are mutually uncorrelated, the covariance matrix is diagonal. 
A detailed discussion regarding the statistical analysis of cosmological models using BAO data is available in reference  \citep{Giostri2012}.

\subsection{CMB shift parameter data:}
The CMB shift parameter, which is related to the position of the first acoustic peak in power spectrum of the temperature anisotropy of the Cosmic Microwave Background (CMB) radiation, is efficient in order to ensure tighter constraints on the model parameters if used in combination with other observational data. The value of CMB shift parameter is not directly measured from CMB observation. The value is estimated from the CMB data along with some fiducial assumption about  the background cosmology. For a spatially flat universe, the CMB shift parameter is defined as
\be
{\mathcal R}=\sqrt{\Omega_{m0}}\int_0^{z_*}\frac{dz}{h(z)},
\label{cmbshift}
\ee 
where $\Omega_{m0}$ is the matter density parameter, $z_*$ is the redshift at photon decoupling and $h(z)=\frac{H(z)}{H_0}$ (where $H_0$ be the present value of Hubble parameter). The $\chi^2_{\tiny CMBShift}$ is defined as 
\be 
\chi^2_{\tiny CMBShift}=\frac{({\mathcal R}_{obs}-{\mathcal R}_{th}(z_*))^2}{\sigma^2},
\ee
where ${\mathcal R}_{obs}$ is the value of the CMB shift parameter, estimated from observation and $\sigma$ is the corresponding uncertainty. In this work, the value of CMB shift parameter estimated from Planck data \citep{WangWang2013b} has been used. As mentioned earlier, it is important to note that for the reconstructed model $\left(\frac{\alpha}{1+\alpha}\right)^{3/n}$ is equivalent to the matter density parameter $\Omega_{m0}$ and hence the $\Omega_{m0}$ of equation (\ref{cmbshift}) has been replaced by $\left(\frac{\alpha}{1+\alpha}\right)^{3/n}$ during the statistical analysis.

\section{Results of Statistical Analysis}

In the present work, $\chi^2$-minimization (which is equivalent to the maximum likelihood analysis) technique has been adopted to estimate the parameter values. For comparison, statistical analysis of $w$CDM dark energy model, consisting two parameter namely the matter density parameter $\Omega_{m0}$ and constant dark energy equation of state parameter $w_{DE}$, has also been carried out with the same technique using the same combinations of the data sets.

\begin{table}
\caption{{\small Results of statistical analysis of the $w_{eff}(z)$ and $w$CDM  model combining OHD, SNe, BAO and CMBShift data}}
\begin{center}
\resizebox{0.495\textwidth}{!}{  
\begin{tabular}{ c |c |c c  } 
  \hline
 \hline
 Model  & $\chi^2_{min}/d.o.f.$ &  Parameters \\ 

 \hline
  $w_{eff}(z)$ & 48.21/52 & $\alpha=0.444\pm0.042$ ; $n=2.907\pm0.136$\\ 
 \hline
  $w$CDM & 48.24/52 & $\Omega_{m0}=0.296\pm0.007$ ;  $w_{DE}=-0.981\pm0.031$\\ 
 \hline
  \hline
\end{tabular}
}
\end{center}
\label{tableweffwCDM}
\end{table}

\begin{figure}
\begin{center}
\includegraphics[angle=0, width=0.22\textwidth]{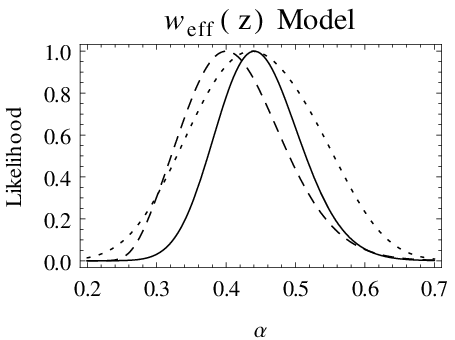}
\includegraphics[angle=0, width=0.22\textwidth]{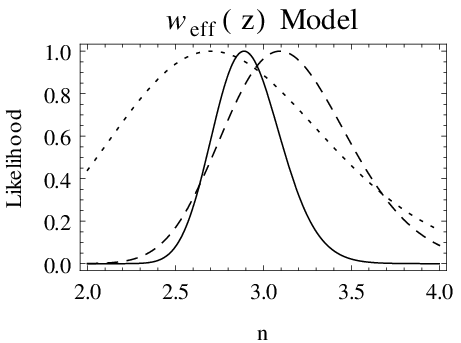}\\
\includegraphics[angle=0, width=0.22\textwidth]{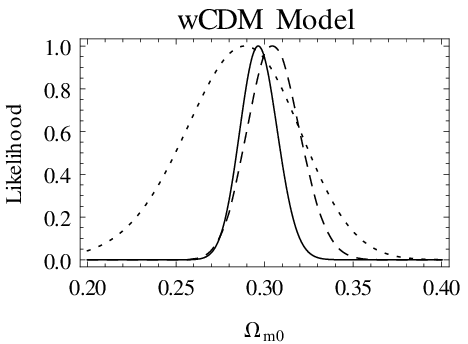}
\includegraphics[angle=0, width=0.22\textwidth]{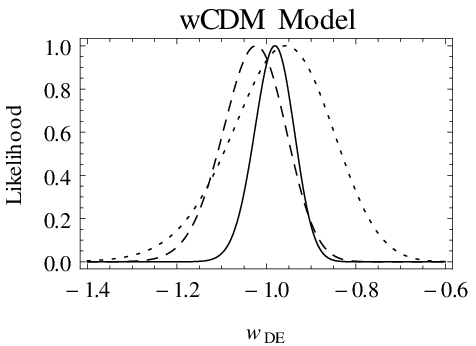}
\end{center}
\caption{{\small Plots  of marginalised likelihood as functions of model parameters. The upper panels show the likelihood for  the $w_{eff}(z)$ model and the lower panels show likelihood of $w$CDM modle obtained from the statistical analysis with different combinations of the data sets. The dotted curves  represent the likelihood obtained for SNe+OHD, the dashed curves are obtained for SNe+OHD+BAO and the solid curves show the likelihood for  SNe+OHD+BAO+CMBShift. }}
\label{wefflikelihood}
\end{figure}

\begin{figure}
\begin{center}
\includegraphics[angle=0, width=0.22\textwidth]{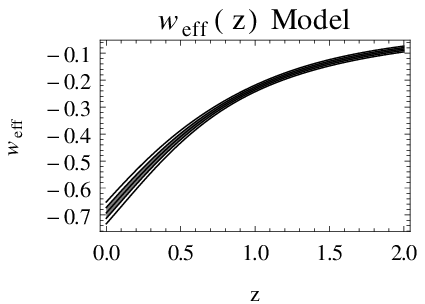}
\includegraphics[angle=0, width=0.22\textwidth]{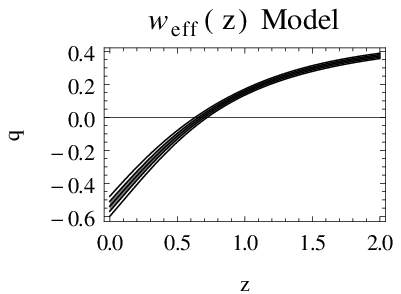}\\
\includegraphics[angle=0, width=0.22\textwidth]{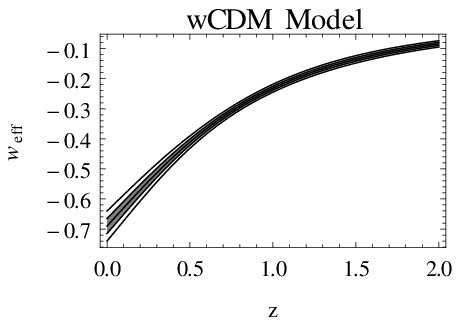}
\includegraphics[angle=0, width=0.22\textwidth]{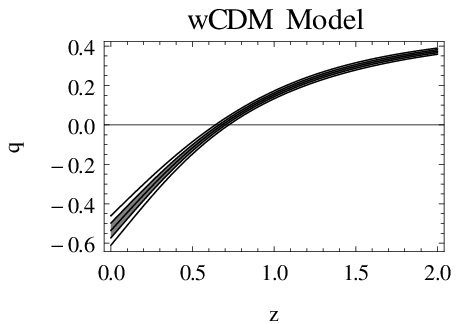}
\end{center}
\caption{{\small Plots of effective equation of state parameter and the deceleration parameter as functions of redshift $z$ for $w_{eff}(z)$ model (upper panels) and $w$CDM model (lower panels). The 1$\sigma$ and  2$\sigma$ confidence regions along with the central black line representing the corresponding the best fit curves obtained from the analysis combining the SNe, OHD, BAO and CMB shift parameter data are presented.}}
\label{weffzqzplot}
\end{figure}
\begin{figure}
\begin{center}
\includegraphics[angle=0, width=0.22\textwidth]{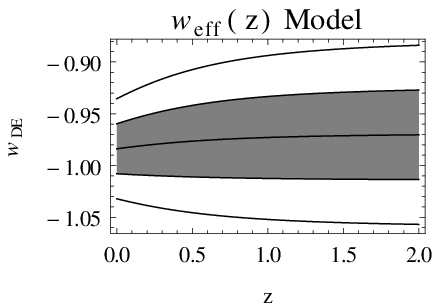}
\includegraphics[angle=0, width=0.22\textwidth]{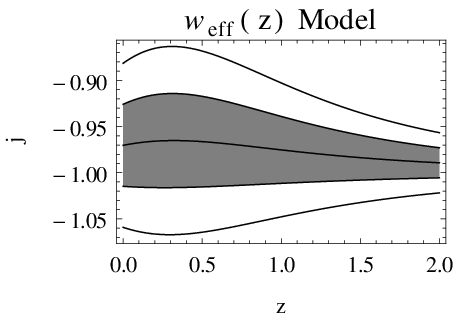}\\
\includegraphics[angle=0, width=0.22\textwidth]{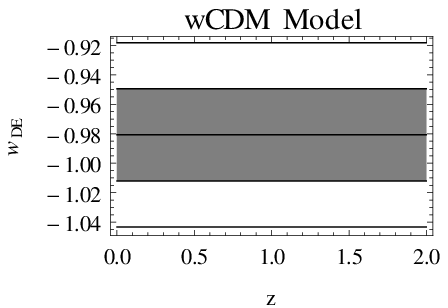}
\includegraphics[angle=0, width=0.22\textwidth]{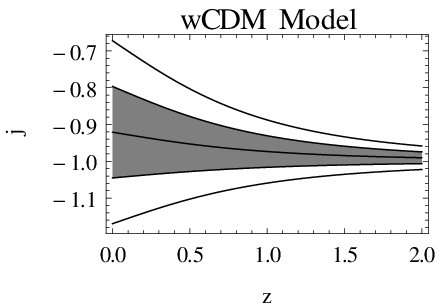}
\end{center}
\caption{{\small Plots of dark energy equation of state parameter ($w_{DE}$) and the cosmological jerk parameter ($j(z)$) parameter as functions of redshift $z$ for $w_{eff}(z)$ model (upper panels) and $w$CDM model (lower panels). The 1$\sigma$ and  2$\sigma$ confidence regions along with the central black line representing the corresponding the best fit curves obtained from the analysis combining the SNe, OHD, BAO and CMB shift parameter data are presented.}}
\label{wDEjzplot}
\end{figure}

Figure \ref{weffcontour} shows the confidence contours on the two dimensional (2D) parameter space for the $w_{eff}(z)$ model for different combinations of the data sets. Similarly figure \ref{wCDMcontour} presents the confidence contours on the 2D parameter space of the $w$CDM models for the same combinations of the data sets. Marginalised likelihoods for the $w_{eff}(z)$ model and $w$CDM model are presented in figure \ref{wefflikelihood}. It is clear from the likelihood plots that the likelihood functions are well fitted to a Gaussian distribution function when all four data sets are take into account. It is also apparent from the confidence contours ( figure \ref{weffcontour} and figure \ref{wCDMcontour}) and the likelihood function plots (figure \ref{wefflikelihood}) that the addition of CMB shift parameter data leads to substantially tighter constraints on the model parameters for both the models. 

Table \ref{tableweffwCDM} contains the results obtained from the statistical analysis combining OHD, SNe, BAO and CMB shift parameter data. The parameters values and the associated 1$\sigma$ uncertainty have been presented along with the reduced $\chi^2$ i.e. $\chi^2_{min}/d.o.f$ of the models. The reduced $\chi^2$ is a measure of the goodness of the fitting. The fitting would be rated to be good if the value of reduced $\chi^2$ be close to one.

\par The model parameter $n$ is important to understand the deviation of the reconstructed model from the $\Lambda$CDM as for $n=3$, the reconstructed model becomes the $\Lambda$CDM. The confidence contours obtained from different combinations of the data sets show that the $\Lambda$CDM is always within 1$\sigma$ confidence regions. The result obtained from the analysis using only the SNe and OHD shows a higher deviation of the best fit value of the parameter $n$ from the corresponding $\Lambda$CDM value than the  results obtained by introducing the BAO and CMB shift parameter data along with SNe and OHD (figure \ref{weffcontour} and the upper right panel of figure \ref{wefflikelihood}). The associated uncertainty obtained from the analysis with SNe+OHD is very large (left panel of figure \ref{weffcontour}) and the constraints become tighter with the addition of other data sets, namely the BAO and CMB shift parameter (middle and right panels of figure \ref{weffcontour}). The best fit  value of $n$ obtained for SNe+OHD is less than 3 (left panel of figure \ref{weffcontour}), for SNe+OHD+BAO it is greater than 3 (middle panel of figure \ref{weffcontour})and for SNe+HOD+BAO+CMBShift, it is slightly less tha 3 (right panel of figure \ref{weffcontour}). So it is apparent that the nature of deviation from $\Lambda$CDM varies according to the combination of data sets used for the analysis. It also deserves mention that the addition of CMB shift parameter data keeps the model in close proximity of $\Lambda$CDM and also ensures much tighter constraints on the parameter values.

\par The deceleration parameter, a dimensionless representation of the second order time derivative of the scale factor, is defined as $q=-\frac{1}{H^2}\frac{\ddot{a}}{a}$. It can also be written in terms of Hubbele parameter and its derivative with respect to redshift as,
\be
q(z)=-1+\frac{1}{2}(1+z)\frac{(h^2)'}{h^2}.
\ee
For the present $w_{eff}$ model, the expression of the deceleration parameter obtained is
\be
q(z)=-1+\frac{3\alpha(1+z)^n}{2(1+\alpha(1+z)^n)}.
\label{presentdecel}
\ee
In figure \ref{weffzqzplot}, the plot of effective equation of state parameter ($w_{eff}$) and the deceleration parameter ($q$) as functions of redshift $z$ for both  the $w_{eff}$ model and $w$CDM model have been presented. The central dark lines represent for the best fit curves and the 1$\sigma$ and 2$\sigma$ confidence region are given from inner to outer part. Figure \ref{weffzqzplot} reveals the fact that the effective equation of state ($w_{eff}$) and deceleration parameter ($q$) evolve in very similar way for both the models. For the proposed model also the deceleration parameter shows a signature flip in between the redshift value $0.6$ to $0.8$, which is well consistent with the analysis of observational data by Farooq and Ratra \citep{FarooqRatra2013}. 

\par Figure \ref{wDEjzplot} shows the plots dark energy energy equation of state parameter $w_{DE}$ and cosmic jerk parameter $j$ for both the models. The jerk parameter $j$, which is the dimensionless representation of the 3rd order time derivative of the scale factor $a(t)$, is defined as
\be
j=-\frac{1}{aH^3}\frac{d^3a}{dt^3}.
\label{jerk}
\ee
It is sometimes defined without the negative sign in the front. In the present work, this convention has been used similar to that of in reference \citep{zz2013,MukherjeeBanerjee2016}. The jerk parameter can also be expressed in terms of Hubble parameter and its derivative with respect to redshift as,
\be
j(z)=-1+(1+z)\frac{(h^2)'}{h^2}-\frac{1}{2}(1+z)^2\frac{(h^2)''}{h^2},
\ee
and for the present model, the expression is
\be
j(z)=-1-\frac{3\alpha(n-3)(1+z)^n}{2(1+\alpha(1+z)^n)}+\frac{3\alpha^2(n-3)(1+z)^{2n}}{2(1+\alpha(1+z)^n)^2}.
\label{presentjerk}
\ee

 The jerk parameter is also important to understand the deviation of the model from $\Lambda$CDM as for a universe with cosmological constant and cold dark matter, the value of jerk parameter is always $-1$. The dark energy equation of state remains almost flat and shows the preference toward the non-phantom nature of dark energy for the reconstructed $w_{eff}$ model (upper left panel of figure \ref{wDEjzplot}). That means its behaviour is very similar to that of $w$CDM model. The plots of $w_{DE}$ and cosmological jerk $j$ show that tighter constraints on their present values are obtained for the reconstructed model than the $w$CDM. For the reconstructed $w_{eff}$ model, plots (upper panels of figure \ref{wDEjzplot}) show that the dark energy equation of state parameter $w_{DE}(z)$ is better constrained at low redshift but the jerk parameter $j(z)$ is better constrained at high redshift. The plots also show that the best fit value of $w_{DE}(z)$ has higher deviation from $-1$ at high redshift and on the other hand the best fit value of $j(z)$ has a higher deviation from the corresponding $\Lambda$CDM value at low redshift. Actually in this case the uncertainty  increases with the increase in deviation of the model from $\Lambda$CDM. It also indicates that the reconstructed model allows a wide variation of the value of $w_{DE}(z)$ at high redshift, but the value of the jerk parameter $j(z)$ is not allowed to have a wide variation at high redshift. The plot of jerk parameter $j(z)$ for the $w$CDM model (lower right panel of figure \ref{wDEjzplot}) also shows a similar behaviour.

\par For statistical comparison of the $w_{eff}(z)$ model to the $w$CDM model, two model selection criterion  have been invoked, the Akaike information criterion (AIC) and the Bayesian Information Criterion (BIC). The Akaike information criterion (AIC) \citep{Akaike}, defined as
\be
AIC=-2\ln{{\mathcal{L}}_{max}}+k.
\ee 
And the Bayesian Information Criterion (BIC) which is based upon the Bayesian Evidence \citep{Schwarz},  defined as
\be
BIC=-2\ln{{\mathcal{L}}_{max}}+k\ln{N},
\ee
where the ${\mathcal{L}}_{max}$ is the maximum likelihood obtained for the model, $k$ is the number of free parameters in that model and $N$ is the number of observational data points used for the statistical analysis. The differences between the AIC of the two models ($\Delta$AIC) and and similarly the difference between the BIC of two models  under consideration ($\Delta$BIC)  are the indicators of consistency between these two models. If the magnitude of $\Delta$BIC or $\Delta$AIC is less than 2, the model under consideration (here $w_{eff}(z)$ model) is strongly favoured by the reference model (here the $w$CDM model). And if it is greater than 10, then the models strongly disfavour each other. Now for the $w_{eff}(z)$ model in comparison with $w$CDM, the  $\Delta$AIC and  $\Delta$BIC vales are
\be
\Delta AIC=\chi^2_{min}(w_{eff}(z))-\chi^2_{min}(wCDM)=-0.03,
\label{deltaBIC}
\ee
and
\be
\Delta BIC=\chi^2_{min}(w_{eff}(z))-\chi^2_{min}(wCDM)=-0.03.
\label{deltaAIC}
\ee
Here $\Delta$AIC and $\Delta$BIC are equal as both the models have two free parameters and the number of data points used of the statistical analysis are same for both the models. The reconstruction of the parametric effective equation of state parameter $w_{eff}$ is highly consistent with the standard dynamical dark energy model.

\section{Representation on $(\MakeLowercase{q_0},\MakeLowercase{j_0})$ parameter space}
For a direct comparison between these two models it would be convenient to look into them through the same parameter space. Now the present value of deceleration parameter $q_0$ and present value of the jerk parameter $j_0$ can be used as the parameters replacing the corresponding model parameters. The $q_0$ and $j_0$ can be obtained from equations (\ref{presentdecel}) and (\ref{presentjerk}) respectively, as 
\be
q_0=-1+\frac{3\alpha}{2(1+\alpha)},
\label{q0}
\ee
and 
\be
j_0=-1-\frac{3\alpha(n-3)}{2(1+\alpha)}+\frac{3\alpha^2(n-3)}{2(1+\alpha)^2}.
\label{j0}
\ee
From these two equations, namely equation (\ref{q0}) and (\ref{j0}), the model parameter $\alpha$ and $n$ can be expressed in terms of $q_0$ and $j_0$.  Substituting those expressions of $\alpha$ and $n$ in equation (\ref{hubbleparam}), $h^2(z)$ for the reconstructed model can be written in terms of $q_0$ and $j_0$ as,
\be
\begin{split}
h^2(z)=\Bigg(\frac{(1-2q_0)}{3}+\frac{2(1+q_0)}{3}\\(1+z)^{\frac{3(1+j_0)+3(1+q_0)(2q_0-1)}{(1+q_0)(2q_0-1)}}\Bigg)^{\frac{(1+q_0)(2q_0-1)}{(1+j_0)+(1+q_0)(2q_0-1)}}.
\label{HubbleWeffq0j0}
\end{split}
\ee
In the same way, the Hubble parameter for $w$CDM model can be expressed in terms of parameter $q_0$ and $j_0$ as
\be  
\begin{split}
h^2(z)=\Bigg(1-\frac{(1-2q_0)^2}{3(1-2q_0)-2(1+j_0)}\Bigg)(1+z)^3\\+\Bigg(\frac{(1-2q_0)^2}{3(1-2q_0)-2(1+j_0)}\Bigg)(1+z)^{\frac{2(1+j_0)}{3-2(1+q_0)}}.
\label{HubblewCDMq0j0}
\end{split}
\ee
A similar statistical analysis has been carried out to estimates the values of the kinematical parameters $q_0$ and $j_0$ for both the models. This type of representation is important for a comparative study of two models. Different model selection criteria or the Bayesian evidence are obviously important to judge the consistency between the models. But the representation on the same parameter space reveals the similarity or difference between the confidence regions of the models under consideration and also shows whether there is some intersection between the confidence contours of the respective models.

\begin{table}
\caption{{\small Results of statistical analysis of the $w_{eff}(z)$ and $wCDM$ model with $q_0$ and $j_0$ as the parameters using the combination of OHD, SNe, BAO and CMBShift data}}
\begin{center}
\resizebox{0.49 \textwidth}{!}{  
\begin{tabular}{ c |c |c c c } 
 \hline
 \hline
 Model  & $\chi^2_{min}/d.o.f.$ & $q_0$ & $j_0$ \\ 

 \hline
  $w_{eff}(z)$ & 49.45/52 & $-0.555\pm0.030$ & $-0.977\pm0.043$\\ 
 \hline
  $w$CDM & 48.24/52 & $-0.535\pm0.037$ &  $-0.940\pm0.094$\\ 
 \hline
 \hline
\end{tabular}
}
\end{center}
\label{tableweffwCDMq0j0}
\end{table}

\begin{figure*}
\begin{center}
\includegraphics[angle=0, width=0.22\textwidth]{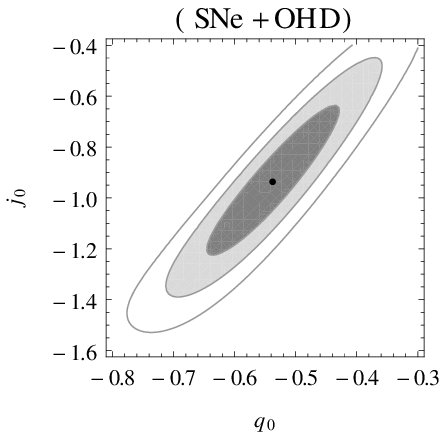}
\includegraphics[angle=0, width=0.22\textwidth]{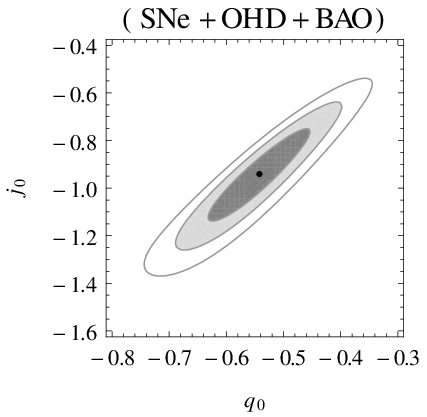}
\includegraphics[angle=0, width=0.22\textwidth]{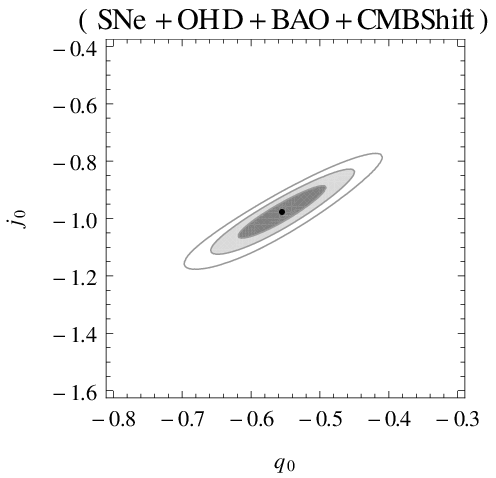}
\end{center}
\caption{{\small Confidence contours on the 2D parameter space ($q_0$,$j_0$) for the $w_{eff}(z)$ model obtained for obtained for different combinations of the data sets. 1$\sigma$, 2$\sigma$ and 3$\sigma$ confidence regions are presented from inner to outer portion and the central black dots represent the corresponding  best fit points. The left panel shows the confidence contours obtained for SNe+OHD, the middle panel shows the confidence contours obtained for SNe+OHD+BAO and the right panel shows confidence contours for SNe+OHD+BAO+CMBShift.}}
\label{weffq0j0Contourplots}
\end{figure*}
\begin{figure*}
\begin{center}
\includegraphics[angle=0, width=0.22\textwidth]{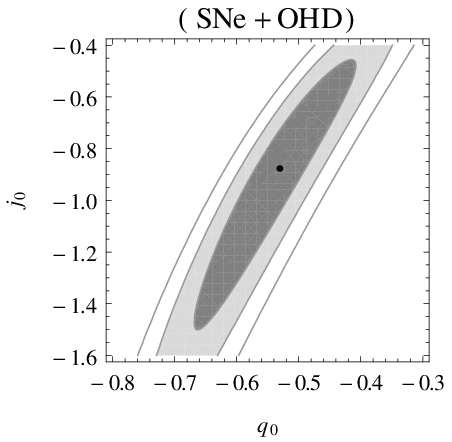}
\includegraphics[angle=0, width=0.22\textwidth]{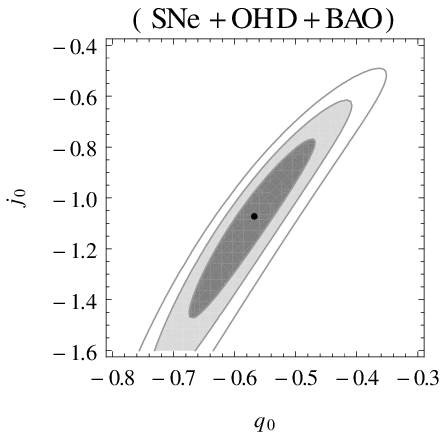}
\includegraphics[angle=0, width=0.22\textwidth]{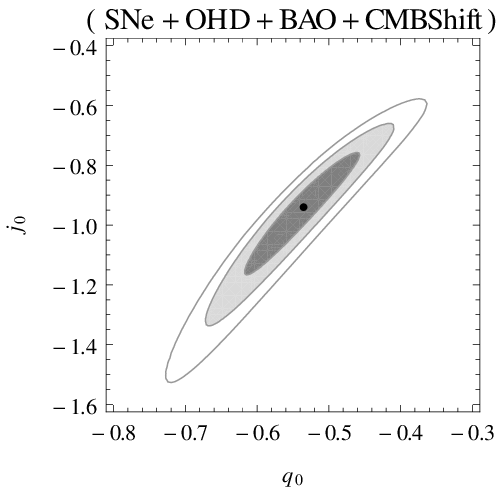}
\end{center}
\caption{{\small Confidence contours on the 2D parameter space ($q_0$,$j_0$) for the $w$CDM model obtained for obtained for different combinations of the data sets. 1$\sigma$, 2$\sigma$ and 3$\sigma$ confidence regions are presented from inner to outer portion and the central black dots represent the corresponding  best fit points. The left panel shows the confidence contours obtained for SNe+OHD, the middle panel shows the confidence contours obtained for SNe+OHD+BAO and the right panel shows confidence contours for SNe+OHD+BAO+CMBShift.}}
\label{wCDMq0j0Contourplots}
\end{figure*}

\begin{figure}
\begin{center}
\includegraphics[angle=0, width=0.22\textwidth]{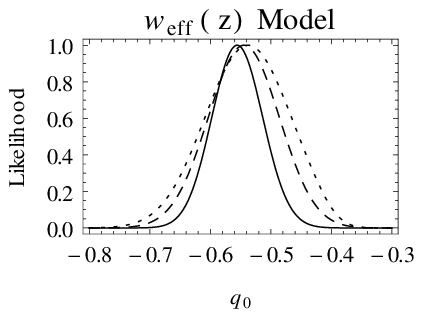}
\includegraphics[angle=0, width=0.22\textwidth]{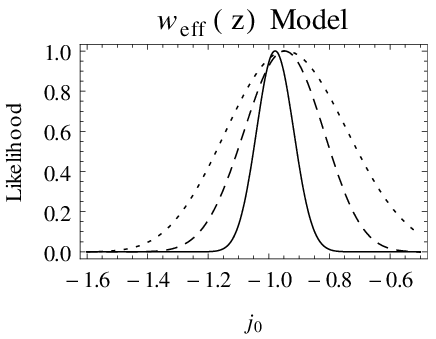}\\
\includegraphics[angle=0, width=0.22\textwidth]{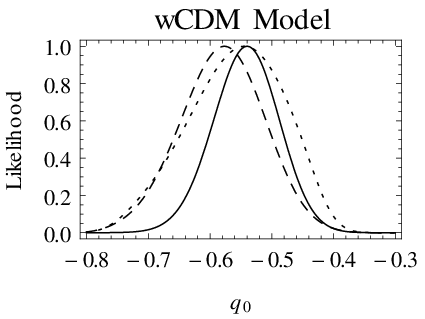}
\includegraphics[angle=0, width=0.22\textwidth]{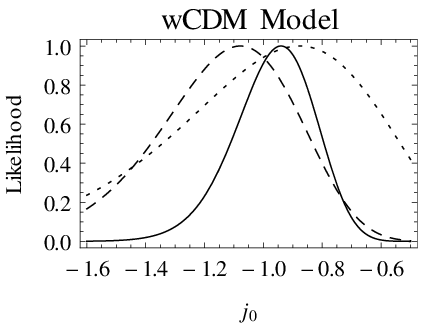}
\end{center}
\caption{{\small The marginalised likelihood as functions of $q_0$ and $j_0$ for the $w_{eff}(z)$ model (upper panels) and for the $w$CDM model (lower panels) obtained for different combinations of the datasets. The dotted curves  represent the likelihood obtained for SNe+OHD, the dashed curves are obtained for SNe+OHD+BAO and the solid curves show the likelihood for  SNe+OHD+BAO+CMBShift. .}}
\label{q0j0Likelihood}
\end{figure}

Table \ref{tableweffwCDMq0j0} presents the results of statistical analysis of the reconstructed $w_{eff}$ model and $w$CDM models respectively obtained from the statistical analysis combining SNe, OHD, BAO and CMB shift parameter data. Figure \ref{weffq0j0Contourplots} and figure \ref{wCDMq0j0Contourplots} show the 2D confidence contours on ($q_0$, $j_0$) parameter space.  Figure \ref{q0j0Likelihood} shows the marginalised likelihood with $q_0$ and $j_0$ as the arguments for both the models.  Though the best fit value of the parameters $q_0$ and $j_0$ obtained for both the models are close enough, the $w_{eff}(z)$ model minimizes the uncertainty of the parameter values. This is also consistent with the results concluded from figure \ref{wDEjzplot}. The representation of the reconstructed model and the $w$CDM model on the ($q_0$,$j_0$) parameter space (figure \ref{weffq0j0Contourplots} and figure \ref{wCDMq0j0Contourplots}) clearly show that the confidence contours of the models are consistent with each other and the correlations between the parameters are very much similar for both the models. It is also clear that substantially tighter constraint on the present value of cosmological jerk parameter ($j_0$) has been achieved for the model reconstructed in the present work.

\section{Conclusion}
 In the present work, the model is built up by considering a parametric form of the effective equation of state parameter. The constraints on the model parameters of the $w$CDM model have also been obtained using the same sets of data. The idea is to draw a direct comparison between these two models. It has been already shown that different model selection criteria clearly indicate the consistency between the $w$CDM and the model reconstructed in the present work.

\par The model parameter $n$ is an indicator of deviation of the model from cosmological constant as for $n=3$ the model exactly mimic the $\Lambda$CDM. Now the value of $n$ obtained from the likelihood analysis is very close to $3$ which indicates the reconstructed $w_{eff}$ is in close proximity of $\Lambda$CDM. As already mentioned that the the contour plots along with the best fit points (figure \ref{weffcontour}) and the likelihood plots (upper right panel of figure \ref{wefflikelihood}) show that the deviations of the model from $\Lambda$CDM vary for different combinations of  the data sets. The addition of CMB shift brings the best fit value of the model parameter $n$ very close to the corresponding $\Lambda$CDM value and much tighter constraints have also been achieved.   In the series expansion expansion of $h^2(z)$, there is a term evolving as $(1+z)^3$. This is equivalent to the matter density and the constant coefficient of this term is the present matter density parameter ($\Omega_{m0}$). For the reconstructed $w_{eff}(z)$ model, the value of the model parameters obtained are $\alpha=0.444\pm0.042$ and $n=2.907\pm0.136$ at 1$\sigma$ confidence level. Consequently the value of the matter density parameter would be $0.296\pm0.011$, which is consistent with the value obtained from the same analysis for $w$CDM model.

\par A recent analysis of $\Lambda$CDM and $w$CDM model by Xia, Li and Zhang \citep{XiaLiZhang20013} using the CMB temperature anisotropy and polarization data along with other non-CMB data estimates the value of the matter density parameter $\Omega_{m0}=0.293\pm0.013$ at 1$\sigma$ confidence level for $\Lambda$CDM and $\Omega_{m0}=0.270\pm0.014$ at 1$\sigma$ confidence level for $w$CDM. Hence the value of the matter density parameter obtained in the present work is very close to the value obtained for $\Lambda$CDM  by Xia, Li and Zhang \citep{XiaLiZhang20013}. A recent analysis by Hazra {\it et al} \citep{HazraDK2015} has presented the analysis of different parameterizations of dark energy using various recent observational data sets. The parameter values obtained are $\Omega_{m0}=0.307_{-0.046}^{+0.041}$, $w_{DE}(z=0)=-1.005_{-0.15}^{+0.17}$ for Chevallier-Polarski-Linder (CPL) parametrization \citep{ChevallierPolarski2001,Linder2003}, $\Omega_{m0}=0.283_{-0.030}^{+0.028}$, $w_{DE}(z=0)=-1.14_{-0.09}^{+0.08}$ for Scherrer and Sen (SS) parameterization \citep{ScherrerSen2008} and $\Omega_{m0}=0.32_{-0.012}^{+0.013}$, $w_{DE}(z=0)=-0.95_{non-phantom}^{+0.007}$ for generalized Chaplygin gas (GCG) model \citep{BentoBertolamiSen2002}. It is clear that the CPL parameterization is in good agreement with the reconstructed $w_{eff}$ model. But the SS parameterization has a preference towards a lower value of the dark energy equation of state though the value of matter density parameter is within 1$\sigma$ confidence region of the reconstructed model. The non-phantom prior assumption of GCG parameterization is in well agreement with the present $w_{eff}$ model, though the present model also allows the phantom behaviour within 1$\sigma$ confidence level. The GCG has a clear preference towards a higher value of the matter density.

\par The plot of deceleration parameter $q(z)$ (upper left panel of figure \ref{weffzqzplot}) shows that the reconstructed model successfully generates the late time acceleration along with the decelerated expansion phase which prevailed before the accelerated expansion phase. The redshift of transition from decelerated to accelerated phase of expansion lies in between the redshift range 0.6 to 0.8 which is consistent with the recent analysis by Farooq and Ratra \citep{FarooqRatra2013}.

\par The equation of state parameter of dark energy achieved for the model presented here remains almost constant. The nature of effective equation of state ($w_{eff}(z)$) and the deceleration parameter $q(z)$ are also very much similar to that of $w$CDM model (figure \ref{weffzqzplot}). Figure \ref{wDEjzplot} presents the plots of $w_{DE}$ and $j(z)$ for the reconstructed $w_{eff}$ model and the $w$CMD model. It is clear from the plots that the reconstructed model puts tighter constraints on the present values of dark energy equation of state parameter ($w_{DE}(z=0)$) and cosmological jerk ($j_0$) than the $w$CDM model. Another interesting point is that the uncertainties associated to the value  of $w_{DE}(z)$ and $j(z)$ vary according to its deviations from the $\Lambda$CDM. A higher deviation of the best fit value from the corresponding $\Lambda$CDM value increases the associated uncertainty. Similar behaviour has also been found in the reconstruction of jerk parameter by Mukherjee and Banerjee \citep{MukherjeeBanerjee2016}. It is also apparently clear from  figure \ref{wDEjzplot} that  the reconstructed $w_{eff}$ model allows a wide variation for the value of dark energy equation of state parameter $w_{DE}(z)$ at high redshift, but the value of the jerk parameter $j(z)$ at high redshift  is not allowed to have a wide variation.    

\par It deserves mention that the systematics of supernova data has been taken into account as it might have its signature on the results. The effect of redshift dependence of colour-luminosity parameter of distance modulus measurement has been discussed by Wang and Wang \citep{WangWang2013a}. There are some recent discussion on the impact of supernova systematics which can also be referred in this context \citep{Rubin2015,ShaferHuterer2015}.

\par As mentioned earlier that the reconstruction of effective or total equation of state is independent of any prior assumption about the nature of dark energy. Though in the present work the reconstructed model allows the matter conservation separately, further generalization with some other ansatz of effective equation of state is possible where interaction between dark energy and dark matter can be taken into account.

\section*{Acknowledgement}
The author would like to thank Professor Narayan Banerjee for guidance and valuable discussions.

\label{SecAppend}
\bibliography{Paper}
\label{lastpage}
\end{document}